\documentstyle[12pt]{article}
\begin{document}

\author{T. P. Singh  \\ %EndAName
Theoretical Astrophysics Group\\Tata Institute of Fundamental Research\\Homi
Bhabha Road, Mumbai 400 005, India\\e-mail: tpsingh@tifrvax.tifr.res.in}
\title{Null Geodesic Expansion in Spherical Gravitational Collapse}
\date{}
\maketitle

\begin{abstract}
\noindent We derive an expression for the expansion of outgoing null
geodesics in spherical dust collapse and compute the limiting value of the
expansion in the approach to singularity formation. An analogous expression
is derived for the spherical collapse of a general form of matter. We argue
on the basis of these results that the covered as well as the naked
singularity solutions arising in spherical dust collapse are stable under
small changes in the equation of state.
\end{abstract}

\newpage\ 

\section{Introduction}

The spherical gravitational collapse of matter obeying an energy condition
has been studied for various equations of state, including dust, perfect
fluids and scalar fields. It is known that both black-hole and naked
singularity solutions can result as the end-state of spherical collapse,
though the issue of the stability of these solutions remains an open one.

The expansion of a congruence of null geodesics, usually denoted by $\theta $%
, can play a useful role in our understanding of the nature of the
singularity that can form in gravitational collapse. With this in mind, we
derive an expression for $\theta $ for spherical dust collapse and also for
spherical collapse of a general form of matter. Using these expressions we
argue for the stability of both the covered and naked singularity solutions
under small perturbations of the equation of state.

\section{Geodesic Expansion in the Dust Model}

In comoving coordinates the Tolman-Bondi metric, which describes spherical
dust collapse, is given by

\begin{equation}
\label{tb}ds^2=dt^2-e^\omega dr^2-R^2d\Omega ^2. 
\end{equation}
The field equations are 
\begin{equation}
\label{fone}\rho =\frac{F^{\prime }}{R^2R^{\prime }} 
\end{equation}
and 
\begin{equation}
\label{ftwo}\dot R^2=\frac FR+f. 
\end{equation}
The energy density $\rho (t,r)$ is the only non-vanishing component of the
energy-momentum tensor. The function $f(r)$ above is defined by the relation 
\begin{equation}
\label{elamb}e^\omega =\frac{R^{\prime 2}}{1+f} 
\end{equation}
and the function $F(r)$ is twice the mass within the comoving coordinate $r$%
. Dot and prime denote differentiation with respect to the comoving
coordinates $t$ and $r$ respectively. We take $\dot R$ to be negative as we
are considering a collapsing cloud.

Consider a congruence of outgoing radial null geodesics in this space-time,
having the tangent vector $(K^t,K^r,0,0)$, where $K^t=dt/dk$ and $K^r=dr/dk$%
. The geodesic expansion $\theta $ is given by

\begin{equation}
\label{thet}\theta =K_{;i}^i=\frac 1{\sqrt{-g}}\frac \partial {\partial
x^i}\left( \sqrt{-g}K^i\right) , 
\end{equation}
which gives

\begin{equation}
\label{thet2}\theta =\frac{\partial K^t}{\partial t}+\frac{\partial K^r}{%
\partial r}+\frac{K^t}{\sqrt{-g}}\left( \sqrt{-g}\right) ^{.}+\frac{K^r}{%
\sqrt{-g}}\left( \sqrt{-g}\right) ^{\prime }. 
\end{equation}
In order to compute the sum 
\begin{equation}
\label{sum}\frac{\partial K^t}{\partial t}+\frac{\partial K^r}{\partial r} 
\end{equation}
we proceed by noting that 
\begin{equation}
\label{kayr}\frac{dK^r}{dk}=\frac{\partial t}{\partial k}\frac{\partial K^r}{%
\partial t}+\frac{\partial r}{\partial k}\frac{\partial K^r}{\partial r}, 
\end{equation}
and similarly 
\begin{equation}
\label{kayt}\frac{dK^t}{dk}=\frac{\partial t}{\partial k}\frac{\partial K^t}{%
\partial t}+\frac{\partial r}{\partial k}\frac{\partial K^t}{\partial r}. 
\end{equation}
Dividing the first of these two relations by $\partial r/\partial k$ and the
second by $\partial t/\partial k$, and after adding the two equations we get 
\begin{equation}
\label{intem}\frac{\partial K^t}{\partial t}+\frac{\partial K^r}{\partial r}%
=\frac 12\left[ \frac 1{K^r}\frac{dK^r}{dk}+\frac 1{K^t}\frac{dK^t}{dk}%
\right] +\frac{\dot \lambda }4K^t-\frac{\lambda ^{\prime }}4K^r. 
\end{equation}
In arriving at this equality we have used the fact that for outgoing radial
null geodesics

\begin{equation}
\label{geo}\frac{K^t}{K^r}=\frac{dt}{dr}=e^{\omega /2}. 
\end{equation}
Further, it follows from the geodesic equation that

\begin{equation}
\label{geo1}\frac{dK^r}{dk}=-\Gamma _{rr}^r\left( K^r\right) ^2-2\Gamma
_{tr}^rK^tK^r, 
\end{equation}
and 
\begin{equation}
\label{geo2}\frac{dK^t}{dk}=-\Gamma _{rr}^t\left( K^r\right) ^2. 
\end{equation}
Together, these relations give the desired expression for $\theta ,$%
\begin{equation}
\label{theeta}\theta (t,r)=\frac{2R^{\prime }}R\left( 1-\sqrt{\frac{f+F/R}{%
1+f}}\right) K^r. 
\end{equation}

Let us now examine the relation that $\theta $ bears with the formation or
otherwise of a naked singularity in spherical dust collapse. As is
well-known, the spherical collapse of a dust cloud results in the formation
of a shell-focussing curvature singularity which lies on the curve $R(t,r)=0$%
. (We assume that the initial density monotonically decreases away from the
center, so that no shell-crossing singularities form). The central
singularity, $R(t_0,0)=0$, is known to be (at least locally) naked for some
initial data and covered for other initial data, whereas the non-central
singularity $R(t,r\neq 0)=0$ is necessarily covered. Consider first the
simpler case of the non-central singularity; here the ratio $F/R$ goes to
infinity as the singularity is approached, while all other quantities in the
expression for $\theta $ are finite. As a result, $\theta $ goes to negative
infinity, which is consistent with the fact that the singularity is covered
and not visible.

The peculiar nature of the central singularity has been analysed in various
previous papers \cite{gref}; in particular we draw attention to our analysis
in \cite{tpj}. There we check for the possible occurrence of a central naked
singularity as follows. We define a quantity $X=R/u$, where $u=r^\alpha $,
and $\alpha $ $>1$ is defined such that $R^{\prime }/r^{\alpha -1}$ is a
unique finite quantity in the limit of approach to $r=0$. Then it is evident
that in the approach to the singularity we can write the limiting value $X_0$
of $X$ as

\begin{equation}
\label{xeq}X_0=\lim _{R\rightarrow 0,r\rightarrow 0}X=\lim \frac Ru=\lim 
\frac{dR}{du}=\lim \frac{R^{\prime }}{\alpha r^{\alpha -1}}\left( 1-\sqrt{%
\frac{f+F/R}{1+f}}\right) . 
\end{equation}
The quantities on the right of the final equality are written as functions
of $X$ and $r$, that is, $X$ is used as a variable instead of $t$. The
occurrence or otherwise of a naked singularity depends on whether or not
this equation has a positive real root. More explicitly, by calculating the
limiting value of $R^{\prime }$ it is shown that the singularity is naked if
the equation 
\begin{equation}
\label{xeq2}X=\frac 1\alpha \left( X+\frac \gamma {\sqrt{X}}\right) \left( 1-%
\sqrt{\frac{f(0)+\Lambda (0)/X}{1+f(0)}}\right) 
\end{equation}
admits a positive root for $X$. The quantity $\gamma $ is a known positive
function of the initial data, and $\Lambda (r)\equiv F/r^\alpha $. It is
seen from the above equation that whenever the singularity is naked the
expression inside the second bracket is positive definite. Further, it is
known that whenever the singularity is covered, this same expression is
negative in the limit.

The connection with the limiting behaviour of $\theta $ is seen as follows.
By comparing the expression for $\theta $ with Eqn.(\ref{xeq2}) above, we
see that the limiting value of $\theta $ is simply $2\alpha /k$ if the
singularity is naked. (We note that near $r=0$, $K^r=dr/dk\approx r/k$).
Hence $\theta $ goes to positive infinity at the naked singularity. On the
other hand, whenever the singularity is covered, $\theta $ goes to negative
infinity. This kind of behaviour of $\theta $ at a covered/naked singularity
is of course to be expected. We merely intend to show that the occurrence of
a central naked singularity is a result of the fact that for certain initial
data the geodesic expansion for outgoing null geodesics continues to be
positive even as the singularity is approached.

We now attempt to ask how the information about the formation or otherwise
of a naked central singularity is contained in the initial distribution for $%
\theta .$ What we do know of course is that for some initial density and
velocity distributions the resulting singularity is naked. We can expect
that somehow this must reflect in the initial profile for $\theta .$ As for
the actual value of $\theta $ at any given point, that is naturally positive
initially, for a congruence of outgoing geodesics - by itself this carries
no information about the nature of the resulting singularity. Hence, we need
to look at the initial distribution of $\theta (r)$ and how it changes from
one point to another. Consider the expression (\ref{theeta}) at a general
time $t$ for the marginally bound case ($f=0$). We recall that the solution
for the area radius is

\begin{equation}
\label{sol}R^{3/2}=r^{3/2}-\frac 32\sqrt{F(r)}t 
\end{equation}
where the scaling $R=r$ at the initial epoch $t=0$ is assumed. Let us also
recall from \cite{tpj} that the series expansion for $F(r)$ near $r=0$ is
written as 
\begin{equation}
\label{fexp}F(r)=F_0r^3+F_qr^{q+3}+... 
\end{equation}
where it is assumed that the first non-vanishing derivative in an expansion
for the density near $r=0$ is the $q$th one. Using this, we can write the
expression (\ref{theeta}) for $\theta $ near $r=0$, to leading order, as 
\begin{equation}
\label{thee2}\theta (t,r)=\frac{2R^{\prime }}R\left( 1-\left( \frac{%
F_0^{3/2}r^3}{1-\frac 32\sqrt{F_0}t-\frac{3F_q}{4F_0}r^qt}\right)
^{1/3}\right) K^r. 
\end{equation}
We concentrate on the behaviour of the expression inside the second bracket,

\begin{equation}
\label{psib}\psi (t,r)\equiv \left( \frac FR\right) ^{3/2}=\frac{F_0^{3/2}r^3%
}{1-\frac 32\sqrt{F_0}t-\frac{3F_q}{4F_0}r^qt}. 
\end{equation}

This expression already contains information as to whether a naked
singularity will result or not, as we now elaborate. The term which carries
information about the inhomogeneity in the distribution is the last term in
the denominator: ( $-\frac{3F_q}{4F_0}r^qt$). The remaining terms in $\psi
(t,r)$ are exactly as they would be for a homogeneous cloud. In the approach
to the time of formation of the central singularity, which is equal to $2/3%
\sqrt{F_0}$, the inhomogeneous term causes $\psi (t,r)$ to go to zero or
infinity depending on whether $q$ is less than three or greater than three.
As a result, $\theta $ goes to a positive limit if $q$ is less than three,
and to a negative limit if $q$ is greater than three. This is consistent
with the fact that the resulting singularity is naked for $q<3$ and covered
for $q>3$. The case $q=3$ is however described satisfactorily by $\psi $
only to an extent: $\theta $ is negative in the limit for $\zeta \equiv
F_3/F_0^{5/2}<-4/3$, whereas it is known that a naked singularity results
only for $\zeta <-25.9904$.

How can we look at the initial data for $\theta $ and decide whether or not
a naked singularity will result? We propose to look at the expression for $%
1/\psi (t,r)$, which is 
\begin{equation}
\label{oneo}\frac 1{\psi (t,r)}=\left( \frac RF\right) ^{3/2}=\frac{1-\frac
32\sqrt{F_0}t}{F_0^{3/2}r^3}-\frac{\frac{3F_q}{4F_0}r^qt}{F_0^{3/2}r^3}. 
\end{equation}
If we look at the inhomogeneous contribution, given by the second term, at
an epoch $t$ just after the start of evolution, we notice that this
contribution diverges at $r=0$, if $q<3$ and goes to zero for $q>3$. (We do
not consider the transition case $q=3$). In fact this continues to be so at
all times $t$. Hence one could suggest that if the inhomogeneous
contribution initially diverges at the center the singularity will be naked,
and if it converges to zero, the singularity will be covered. Physically
this means that at the initial epoch, the inhomogeneity can cause the
geodesic expansion to either decrease or increase as one moves away from the
center. The former case results in a naked singularity, and the latter in a
covered singularity. The same set of arguments also hold for the
non-marginally bound case ($f\neq 0$).

\section{Geodesic expansion in the general spherical case}

We now show that the expression for $\theta $ in the case of general
spherical collapse can be cast in exactly the same form as for dust
collapse. To begin with, we write the Einstein equations for the general
spherical case in comoving coordinates. The metric, in comoving coordinates,
is 
\begin{equation}
\label{gensph}ds^2=e^\sigma dt^2-e^\omega dr^2-R^2d\Omega ^2 
\end{equation}
and the energy-momentum tensor is $T_k^i=diag(\rho ,p_r,p_T,p_T)$. The field
equations for this system are

\begin{equation}
\label{mprime}\rho =\frac{F^{\prime }}{R^2R^{\prime }}, 
\end{equation}

\begin{equation}
\label{mdot}\dot F=-p_rR^2\dot R, 
\end{equation}

\begin{equation}
\label{sigpri}\sigma ^{\prime }=-\frac{2p_r^{\prime }}{\rho +p_r}+\frac{%
4R^{\prime }}{R(\rho +p_r)}(p_T-p_r), 
\end{equation}

\begin{equation}
\label{omedot}\dot \omega =-\frac{2\dot \rho }{\rho +p_r}-\frac{4\dot R(\rho
+p_T)}{R(\rho +p_r)}, 
\end{equation}
and

\begin{equation}
\label{energy}e^{-\sigma }\dot R^2=\frac FR+f. 
\end{equation}
The function $F(t,r)$ is equal to twice the mass inside the comoving
coordinate $r$, and as in the dust case, the function $f(t,r)$ is defined by
the relation (\ref{elamb}). The difference from the dust case is that now
the functions $F$ and $f$ depend on time as well. (It is only for the dust
equation of state that both these functions are time-independent). When one
considers the special case of dust, the above five equations behave as
follows. Eqn. (\ref{mprime}) holds as such, while (\ref{energy}) reduces to (%
\ref{ftwo}). Equations (\ref{mdot}) and (\ref{sigpri}) are trivially
satisfied, whereas (\ref{omedot}) reduces to an identity.

The calculation of $\theta $ proceeds exactly as in the case of dust, and we
get again the expression in Eqn. (\ref{thet2}). In order to calculate the
sum (\ref{sum}) we use the relations (\ref{kayr}) and (\ref{kayt}) and we
note that now for outgoing null geodesics we have 
\begin{equation}
\label{geog}\frac{K^t}{K^r}=\frac{dt}{dr}=e^{(\omega -\sigma )/2}. 
\end{equation}
Hence we get, by proceeding as in the dust case, 
\begin{equation}
\label{intem2}\frac{\partial K^t}{\partial t}+\frac{\partial K^r}{\partial r}%
=\frac 12\left[ \frac 1{K^r}\frac{dK^r}{dk}+\frac 1{K^t}\frac{dK^t}{dk}%
\right] +\frac{\dot \lambda -\dot \nu }4K^t-\frac{\lambda ^{\prime }-\nu
^{\prime }}4K^r. 
\end{equation}
By substituting for $dK^r/dk$ and $dK^t/dk$ from the geodesic equation we
again find that $\theta $ is given by 
\begin{equation}
\label{thea}\theta (t,r)=\frac{2R^{\prime }}R\left( 1-\sqrt{\frac{f+F/R}{1+f}%
}\right) K^r, 
\end{equation}
which is the same expression as in the dust case, except that now $F$ and $f$
are functions of $t$ as well as $r$. We have also utilised the field
equation (\ref{energy}) which in fact is the only field equation used in
writing the above expression for $\theta $.

\section{Discussion}

We note that $\theta $ for dust collapse has a peculiar behaviour in the 
limit of approach to
the singularity curve. As regards the points other than the center (i.e.
those with $r\neq 0$), they first get trapped (i.e. $\theta $ becomes zero),
and then singular. This is consistent with the singularity theorems - the
formation of a trapped surface at a given $r$ is followed by the formation
of a singularity for this value of $r$. However, $r=0$ is a very special
point: the apparent horizon curve begins to form at $r=0$ simultaneously
with the occurrence of the singularity. As a result, it would appear that
one could not directly appeal to a singularity theorem to predict the
formation of the central singularity, even though an explicit calculation
shows that a singularity does form at $r=0$. Similarly, one cannot predict a
priori that the central singularity will necessarily be covered, and
explicit calculation actually shows it to be otherwise.

We would like to argue here, judging by the expression for $\theta $, that
both the black-hole and naked singularity solutions arising in spherical
dust collapse are stable under sufficiently small changes of equation of
state, for a fixed initial data. This is expected to hold for all initial
data, except near the transition region ( i.e. the case $q=3$) in the above
discussion, and the reasoning is as follows. Consider an initial density and
velocity distribution in spherical dust collapse which results in a covered
singularity. As discussed above, this covered singularity results because
for this initial data, the ratio $F/R$ goes to positive infinity in the
limit of approach to the singularity. We have noticed, in Eqn. (\ref{oneo}),
that the initial data already contains the information that $F/R$ will
behave in this particular manner in the approach to the singularity, thereby
making it a covered singularity. Now, if we keep the initial density and
velocity distribution fixed, and make a sufficiently small change in the
equation of state, the initial behaviour of $F/R$ will be arbitrarily close
to that in the dust case. This is to be expected from the stability of
solutions of Einstein equations under a small change in the equation of
state. Hence the evolution will still be such that $F/R$ will again go to
positive infinity as the singularity is approached, which makes $\theta $
negative, and hence the singularity continues to be covered. (The function $%
f(r)$ plays an insignificant role in this argument: one only has to make the
plausible demand that $f(0)$ be finite during the evolution). A similar
argument ensures that the dust naked singularity solutions will be stable
under perturbation of the equation of state (in this case $F/R$ goes to zero
in the limit and $\theta $ is positive). It is only in the transition region
($q=3$) that both the naked and covered solutions could be unstable: one
kind of solution could go into the other kind.

From an argument of this nature it also follows that if for any given
equation of state some initial data lead to a covered (naked) singularity,
the nature of the singularity will be left unchanged if the equation of
state is perturbed while keeping the initial conditions fixed. The same
conclusion has also been arrived at earlier in \cite{dj} by independent
means. It appears as if both the covered and naked solutions arising in
spherical collapse must be treated on the same footing, in so far as
stability under change of equation of state is concerned.


\begin{thebibliography}{9}
\bibitem{gref}  D. M. Eardley and L. Smarr, Phys. Rev. D {\bf 19} (1979)
2239; D. Christodoulou, Commun. Math. Phys. {\bf 93} (1984) 171; R. P. A. C.
Newman, Class. Quantum Grav. {\bf 3} (1986) 527; I. H. Dwivedi and P. S.
Joshi, Class. Quantum Grav. {\bf 9} (1992) L69; P. S. Joshi and I. H.
Dwivedi, Phys. Rev. {\bf D47} (1993) 5357. 

\bibitem{tpj}  T. P. Singh and P. S. Joshi, Class. Quantum Gravity {\bf 13 }%
(1996) 559.

\bibitem{dj}  I. H. Dwivedi and P. S. Joshi, Commun. Math. Phys. {\bf \ 166}
(1994) 117.
\end{thebibliography}
\end{document}